\documentclass[%
 reprint,
superscriptaddress,
 amsmath,amssymb,
 aps,
prl,
]{revtex4-1}

\usepackage{pdfpages} 
\usepackage{pgffor} 

\makeatletter
\AtBeginDocument{\let\LS@rot\@undefined}
\makeatother

\newif\ifarXiv
\arXivtrue 
\usepackage{natbib}
\usepackage{graphicx}
\usepackage{epstopdf}
\usepackage{dcolumn}
\usepackage{bm}
\usepackage{gensymb}
\usepackage{upgreek}
\usepackage{hyperref}
\usepackage{graphicx}
\usepackage{epstopdf}
\usepackage{dcolumn}
\usepackage{bm}
\usepackage{gensymb}
\usepackage{upgreek}
\usepackage{hyperref}
\usepackage{lineno}
\usepackage{xr}
\makeatletter
\newcommand*{\addFileDependency}[1]{
  \typeout{(#1)}
  \@addtofilelist{#1}
  \IfFileExists{#1}{}{\typeout{No file #1.}}
}
\makeatother

\begin{document}

\title{Ultrafast optical observation of spin-pumping induced dynamic exchange coupling in ferromagnetic semiconductor/metal bilayer}

\author{X. Liu}
\thanks{These authors contributed equally to the work.}
\affiliation{Department of Applied Science, The College of William and Mary, Williamsburg, Virginia, 23187, USA}

\author{P. Liu}
\thanks{These authors contributed equally to the work.}
\affiliation{Department of Applied Science, The College of William and Mary, Williamsburg, Virginia, 23187, USA}

\author{H. C. Yuan}
\affiliation{Key Laboratory of Micro and Nano Photonic Structures (Ministry of Education), Shanghai Ultra-precision Optical Manufacturing Engineering Research Center, Department of Optical Science and Engineering, Fudan University, Shanghai, 200433, China}

\author{J. Y. Shi}
\affiliation{Key Laboratory of Micro and Nano Photonic Structures (Ministry of Education), Shanghai Ultra-precision Optical Manufacturing Engineering Research Center, Department of Optical Science and Engineering, Fudan University, Shanghai, 200433, China}
\author{H. L. Wang}
\author{S. H. Nie}
\affiliation{State Key Laboratory of Supperlattices and Microstructures, Institute of Semiconductors, Chinese Academy of Sciences, Beijing, 100083, China}
\author{F. Jin}
\author{Z. Zheng}
\affiliation{Key Laboratory of Micro and Nano Photonic Structures (Ministry of Education), Shanghai Ultra-precision Optical Manufacturing Engineering Research Center, Department of Optical Science and Engineering, Fudan University, Shanghai, 200433, China}
\author{X. Z. Yu}
\author{J. H. Zhao}
\thanks{email: jhzhao@red.semi.ac.cn}
\affiliation{State Key Laboratory of Supperlattices and Microstructures, Institute of Semiconductors, Chinese Academy of Sciences, Beijing, 100083, China}
\author{H. B. Zhao}
\thanks{email: hbzhao@fudan.edu.cn}
\affiliation{Key Laboratory of Micro and Nano Photonic Structures (Ministry of Education), Shanghai Ultra-precision Optical Manufacturing Engineering Research Center, Department of Optical Science and Engineering, Fudan University, Shanghai, 200433, China}
\author{G. Lüpke}
\thanks{email: gxluep@wm.edu}
\affiliation{Department of Applied Science, The College of William and Mary, Williamsburg, Virginia, 23187, USA}

\date{\today}

\begin{abstract}
Spin angular momentum transfer in magnetic bilayers offers the possibility of ultrafast and low-loss operation for next-generation spintronic devices. We report the field- and temperature- dependent measurements on the magnetization precessions in Co$_2$FeAl/(Ga,Mn)As by time-resolved magneto-optical Kerr effect (TRMOKE). Analysis of the effective Gilbert damping and phase shift indicates a clear signature of an enhanced dynamic exchange coupling between the two ferromagnetic (FM) layers due to the reinforced spin pumping at resonance. The temperature dependence of the dynamic exchange-coupling reveals a primary contribution from the ferromagnetism in (Ga,Mn)As. 
\end{abstract}

\maketitle

\section{\label{sec:Intro}Introduction}

There has been growing interest in the ultrafast optical manipulation of magnetic dynamics in ferromagnetic heterostructures due to its potential applications in advanced functional spintronic devices. The spin-pumping (SP) effect, in which a spin-precessing ferromagnetic layer transfers its angular momentum into another layer by a chargeless spin current, brings a new mechanism for spin controlling and hence plays an important role in the design of future spintronic devices.\cite{1} Since Heinrich, B. et al first reported the spin-pumping effect as increased damping of the source layer in ferromagnetic resonance (FMR) experiments \cite{2} and a few FMR experiments on SP effect have been performed on transition-metal multilayers \cite{3,4,5,6}, topological insulators \cite{7,8} and semiconductors \cite{9,10,11}. In addition, Danilov A. et al. demonstrated that the mutual SP effect modifies the precession dynamics in a pseudo spin-valve where magnetization precessions are excited simultaneously in two FM layers by femtosecond laser pulses.\cite{12} However, no SP effect has ever been observed yet for the heterostructure of a Heusler alloy and a ferromagnetic (FM) semiconductor. Importantly, the hard and soft ferromagnetic phases in such materials can potentially exhibit a dynamic exchange coupling that is completely independent of the static exchange coupling due to spin pumping. This could offer a possibility of ultrafast low-power control of spin current for next-generation spintronic devices.

In this study, we investigate the magnetization precession dynamics of the Heusler alloy Co$_2$FeAl/FM semiconductor (Ga,Mn)As heterostructure as a function of applied field and temperature by time-resolved magneto-optical Kerr effect (TRMOKE). Analysis of the field-dependent effective Gilbert damping indicates a clear signature of the enhanced dynamic exchange coupling between the two FM layers due to a reinforced spin pumping. In addition, curvatures of the phase shift as a function of applied field elucidate the dynamic exchange-coupling model where the counter-precessing precessions are damped significantly at the resonant frequency of the two FM layers. The magnetization precession in the Co$_2$FeAl layer transfers a pure spin current directly into the ferromagnetic semiconductor (Ga,Mn)As layer without a nonmagnetic metal spacer. On the other hand, the temperature-dependent results manifest a strong contribution from the ferromagnetism of (Ga,Mn)As to the dynamic exchange-coupling effect. These results provide valuable insight into the topic of dynamic exchange coupling and the detection of spin current. Furthermore, they suggest a new pathway of ultrafast spin manipulation in metal/semiconductor bilayer systems at low power and therefore promote the development and design of future spintronic devices.

\section{Results and Discussion}

\begin{figure}
    \centering
    \advance\leftskip-0.8cm
    \advance\rightskip-0cm
    \includegraphics[width=0.52\textwidth]{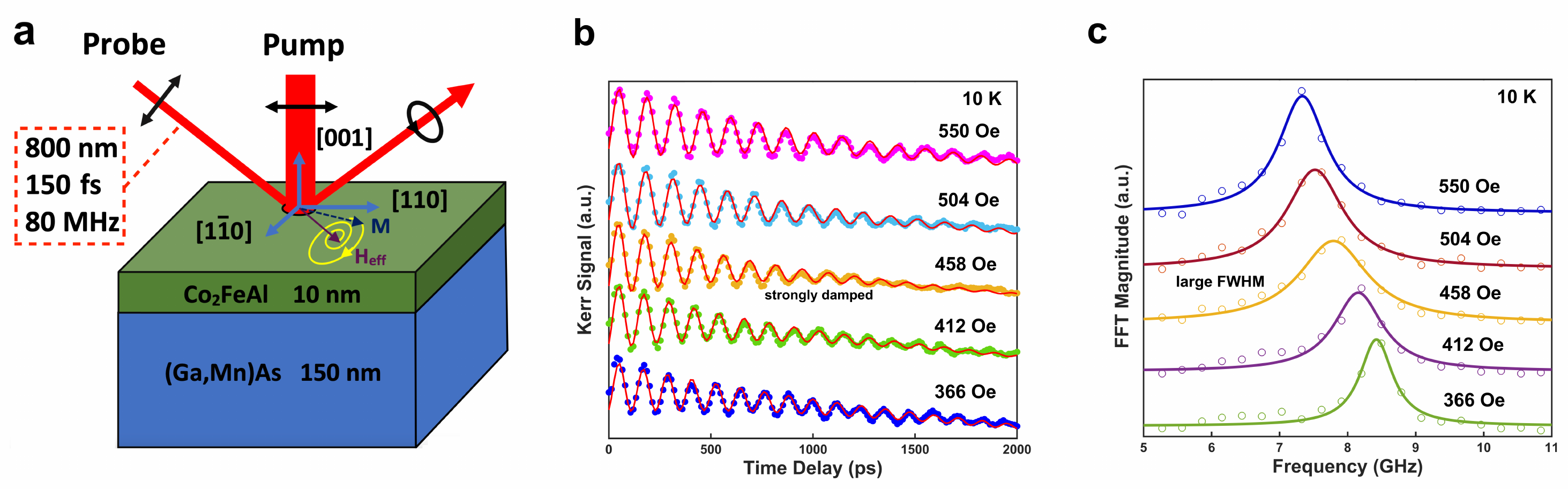}
    \caption{(a) Schematic of TRMOKE measurement geometry, depicting the structure of the sample and the magnetization M precessing around the effective field Heff in Co$_2$FeAl/(Ga,Mn)As bilayer in a canted magnetization configuration with H applied along hard axis [1-10]. (b) TRMOKE data from Co$_2$FeAl/(Ga,Mn)As bilayer under the different intensity of the applied field. (c) FFT analysis for magnetization precession frequency under the different intensities of the applied field, where the solid lines represent the FFT peaks fitted by Lorentz functions.
}
    \label{fig:Fig1}
\end{figure}

The Co$_2$FeAl/(Ga,Mn)As bilayer sample is grown on GaAs (001) substrates by molecular-beam epitaxy (MBE). The thickness of Co2FeAl and Ga1-xMnxAs (x=0.07) layer is 10 nm and 150 nm, respectively. The sample is capped with 2-nm thick Al layer to avoid oxidation and contamination. The hard FM Co$_2$FeAl exhibits an in-plane uniaxial magnetic anisotropy with an easy axis along the [110] direction (Fig.~\ref{fig:Fig1}(a)), whereas the easy axis of the soft FM (Ga,Mn)As is along the [1-10] direction below its Curie temperature $T_c$ = 50 K.\cite{13} Reflection high-energy electron diffraction (RHEED) patterns, high-resolution double-crystal x-ray diffraction (DCXRD) measurements, and high-resolution cross-sectional transmission electron microscopy (HRTEM) reveal high-quality, single-crystalline, epitaxial growth of the Co2FeAl and (Ga,Mn)As thin films.\cite{13} At low temperatures ($T < T_c$), a ferromagnetic alignment of local Mn moments in the (Ga,Mn)As layer is expected, whereas at high temperatures ($T > T_c$) the Mn ions extending a few nanometers from the interface remain spin-polarized due to the ferromagnetic proximity effect.\cite{13} 

Figure~\ref{fig:Fig1}(a) shows the experimental geometry of TRMOKE measurements. Field-dependent TRMOKE measurements are performed on the Co$_2$FeAl/(Ga,Mn)As bilayer sample from 7 K to 300 K utilizing 100-fs pump and probe pulses at 800 nm wavelength with a repetition rate of 80 MHz. The external magnetic field is set along the easy-hard axis [100] of the Co$_2$FeAl layer and the pump fluence is set at $I$ = 5 $\mu$J/cm$^2$. The probe pulses utilize the balanced detection technique with a half-wave plate and Wollaston prism to investigate the transient magnetic state change along longitudinal and polar directions. Figure 1b displays the TRMOKE data at 10 K with an in-plane magnetic field scanning from 366 Oe to 550 Oe. The precession signals can be well fitted by a damped-harmonic function with a linear background: $\theta_k = a_0+b_0t+A\textrm{ exp}\left ( -t/\tau\right)\textrm{ sin}\left(2\pi ft+\varphi_0\right)$ , where $a_0+b_0 t$ represents the linear background, $A$ is the precession amplitude, $\tau$ is the relaxation time, f is the precession frequency and $\varphi_0$ is the phase. The magnetization precession decays with different relaxation times, with the fastest decay at 458 Oe. This indicates that a dynamic exchange coupling may occur between the magnetization precession in the Co$_2$FeAl layer and the (Ga,Mn)As layer. Here, only one frequency can be extracted from FFT analyses, as seen in Fig.~\ref{fig:Fig1}(c), coherent spin precession of (Ga,Mn)As decays fast and thus vanishes very shortly.

The magnetization precession in the Co$_2$FeAl/(Ga,Mn)As bilayer system is described by the following modified Landau-Lifshitz-Gilbert(LLG) equation with an additional spin-torque term:

\small
\begin{equation*}
    \frac{d \mathbf{M}}{d t}=-\gamma \mathbf{M} \times \mathbf{H_{eff}}+\alpha_0 \mathbf{M}\times \frac{d \mathbf{M}}{d t}+\alpha_{sp} \left ( \mathbf{M} \times \frac{d \mathbf{M}}{d t}-\mathbf{M}^\prime \times \frac{d \mathbf{M}^\prime}{d t}\right )
    \label{eq:eq1}
\end{equation*}
\normalsize
where $\mathbf{M}$ is the magnetization direction of the Co$_2$FeAl layer, $\gamma$ is the gyromagnetic ratio, $\alpha_0$ is the intrinsic Gilbert damping constant, and $\mathbf{H_{eff}}$ is the effective magnetic field in the Co$_2$FeAl layer including the external magnetic field, the demagnetization field, the anisotropy field, and the exchange-coupling field. The last term describes the spin torque which acts on both layers as a bidirectional effect, in which $\alpha_{sp}$ represents the contribution of spin pumping to the damping and $\mathbf{M}^\prime$ denotes the magnetization of (Ga,Mn)As. Then, the effective Gilbert damping can be obtained from the relaxation time $\tau$, using \cite{14,15} 

\begin{equation*}
    \alpha = 2/\left ( \tau\gamma \left (H_a+H_b \right )\right )
    \label{eq:eq1}
\end{equation*}

where $H_a$ and $H_b$ are determined in our previous analysis, which includes the out-of-plane, in-plane uniaxial, crystalline cubic, unidirectional and rotatable magnetic anisotropies.\cite{15} 

\begin{figure}
    \centering
    \advance\leftskip-0.8cm
    \advance\rightskip-0cm
    \includegraphics[width=0.52\textwidth]{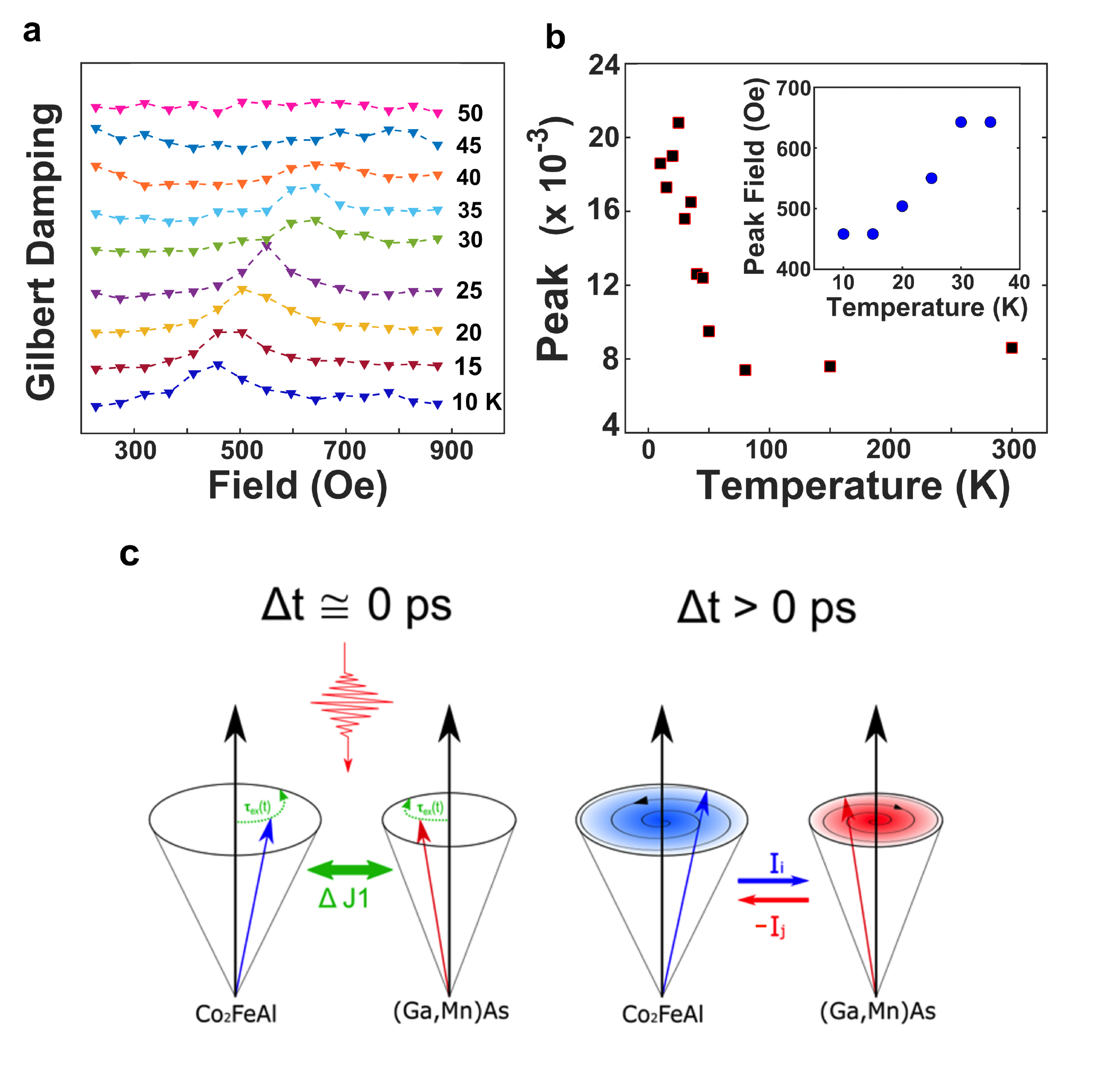}
    \caption{(a) Effective Gilbert damping constant as a function of externally applied field at different temperatures from 10 K to 50 K. (b) Red rectangular solids denote the peak (or strongest if not peak) Gilbert damping from 10 K to 50 K. Blue circle solids are the corresponding fields as a function of temperature. (c) Illustration of photo-excited exchange-coupling torque and spin-pumping generated dynamic exchange-coupling mode for damping.
}
    \label{fig:Fig2}
\end{figure}

Figure~\ref{fig:Fig2}(a) shows the temperature-dependent Gilbert damping as a function of the external field. At $T$ = 10 K, the damping of the magnetization precession is most pronounced with an external field $H$ = 458 Oe. Below the Curie temperature ($T_c$ = 50 K) of (Ga,Mn)As, the damping peak first shifts from 450 Oe to 650 Oe with the temperature increasing from 10 K to 35 K, as summarized in Fig.~\ref{fig:Fig2}(b) insert, and then gradually becomes inconspicuous and finally disappears at $T = T_c$. Such a temperature dependency clearly shows that the damping peak only exists when the ferromagnetism of (Ga,Mn)As is well-developed. Meanwhile, the strongest Gilbert damping extracted across all the fields as a function of temperature (Fig.~\ref{fig:Fig2}(b)) shows a transition temperature close to $T_c$. This manifests the crucial role of spontaneous (Ga,Mn)As magnetization in the damping of magnetization precession of Co$_2$FeAl. 

The ultrafast pump excitation causes a transient enhancement of exchange coupling,\cite{15} which induces a dynamic exchange-coupling torque acting on both Fe(Co) spins and Mn spins. In such a case, as shown in Fig.~\ref{fig:Fig2}(c), the magnetizations of both FM layers are suddenly pulled towards each other and start to precess with opposite angular momentum along their own equilibrium directions. At the resonance, i.e., $f_{CFA} = f_{GMA}$, the precessing magnetization of Co$_2$FeAl “pumps” a spin current $I_i$ directly into the (Ga,Mn)As layer, which exerts a torque onto the (Ga,Mn)As magnetization and thereby counteracts its precession. Meanwhile, this spin current $I_i$ carries an outflow of angular momentum from the Co$_2$FeAl layer and leads to damping to its magnetization precession. In other words, the spin current reinforces the $M$ damping for both FM layers at the resonance. Technically, there should also be a spin current $I_j$ injecting into the Co$_2$FeAl layer from the magnetization precession of (Ga,Mn)As. \cite{2} However, such a spin current should be much smaller than that from the Co$_2$FeAl layer. 

\begin{figure}
    \centering
    \advance\leftskip-0.4cm
    \advance\rightskip-0cm
    \includegraphics[width=0.52\textwidth]{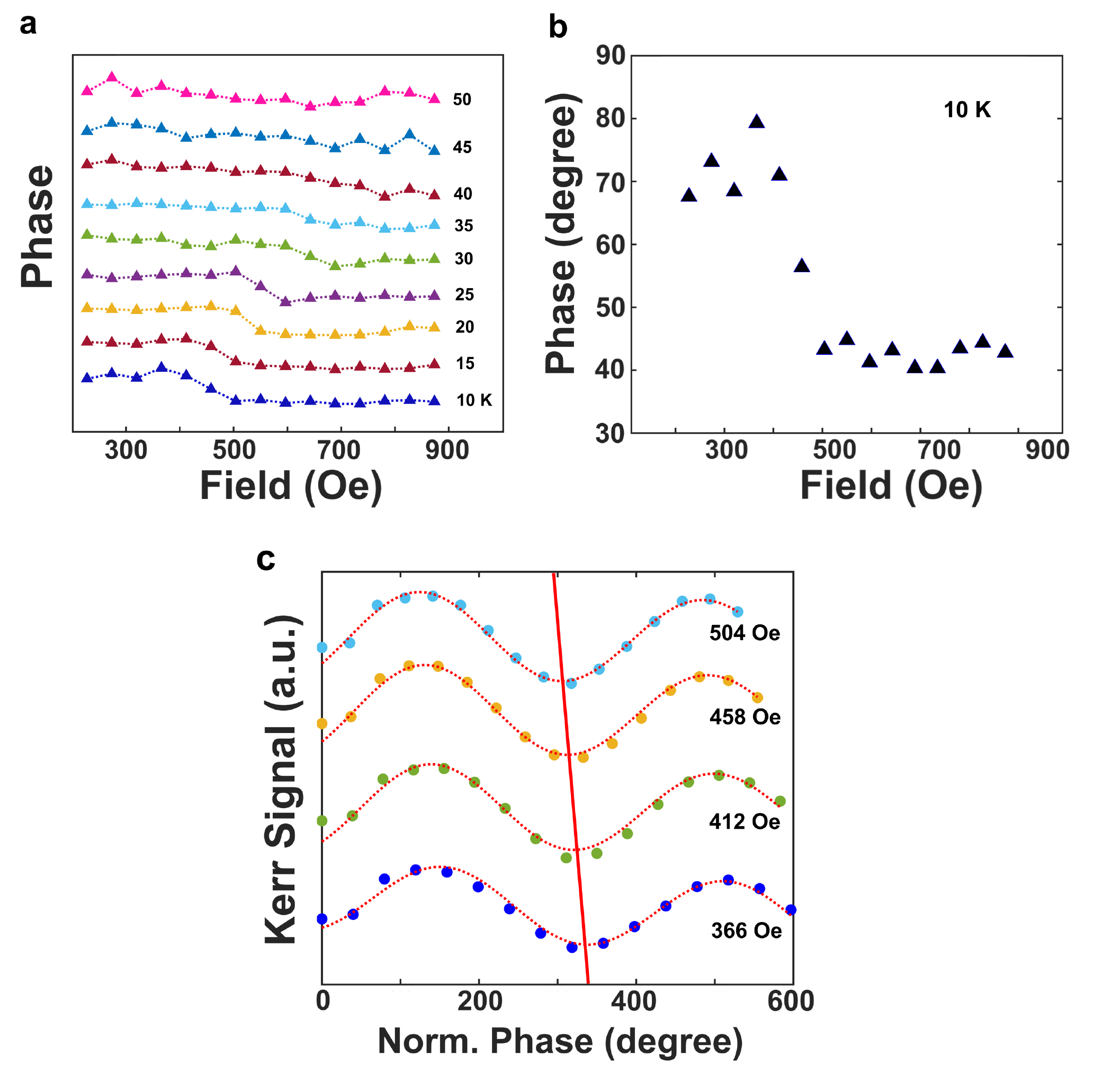}
    \caption{Phase of magnetization precession as a function of externally applied field at different temperatures from 10 K to 50 K (a), and zoomed specifically at 10 K (b). (c) TRMOKE data as a function of normalized precession phase under different applied fields. The red straight line is a guideline of the phase shift. 
}
    \label{fig:Fig3}
\end{figure}

In addition to the discussion on damping, the dynamic exchange coupling between the two FM layers can also be evinced by the field-dependency of the precession phase. Fig.~\ref{fig:Fig3}(a) shows that from 10 K to 35 K, the phase drops down dramatically around certain field windows that correspond to the Gilbert damping peaks, which move to higher field ranges as temperature increases. When $T >$ 30 K, the dramatic phase shift becomes less contrastive and then completely disappears when $T = T_c$. We notice that similar features of phase-shifting are reported in the FMR experiments on similar hard/soft FM systems.\cite{16,17} The observed 30° -- 40° phase shift at 10 K, as shown in Fig.~\ref{fig:Fig3}(c), is comparable with those of the dynamic exchange-coupling spin-valve structures.\cite{7,16,17,18,19}

\section{\label{sec:Conclusion}Conclusion}

In summary, we have studied the dynamics of the magnetization precession of Co$_2$FeAl/(Ga,Mn)As heterostructure as a function of applied field and temperature. The pronounced peaks in effective Gilbert damping of Co$_2$FeAl magnetization precession reveal the enhanced dynamic exchange coupling between the two FM layers due to spin pumping. The corresponding field-dependency of phase shift, which corresponds to that of the Gilbert damping, confirms the counter-precessing exchange-coupled model where both precessions are damped significantly at the resonance. In addition, the temperature-dependent results manifest a strong contribution from the ferromagnetism in (Ga,Mn)As to the dynamic exchange-coupling effect. These results provide valuable insight into the topic of dynamic exchange coupling and the detection of spin current. Mover, they suggest a novel route of ultrafast low-power spin manipulation in metal/semiconductor bilayer system and hence promote the research of the future spintronic devices.

\section{\label{sec:Methods}Methods}

\textbf{MOKE experiments.} The magnetization of the exchange-coupled Co$_{2}$FeAl/(Ga,Mn)As bilayer is measured using a longitudinal MOKE setup. The sample is illuminated with p-polarized light and the reflected s-polarized light is detected with a photodiode. The magnetic field is applied along the in-plane [110] or [-110] crystallographic directions. The measurements are conducted from 5 K to above room temperature. 

\textbf{TRMOKE experiments.} For the pump-probe TRMOKE measurements, a Ti:sapphire oscillator laser system is employed, which produces 150-fs pulses at 800-nm wavelength with a repetition rate of 80 MHz. The probe(pump) fluence is fixed at $\sim$0.5(5) $\mu$J/cm$^{2}$. The probe pulses ($\lambda=800$ nm) use the balanced detection approach with a half-wave plate and Wollaston prism to investigate the transient magnetic state change along longitudinal and polar directions. The measurements are conducted from 5 K to above room temperature.

\section{Data Availability}
The data that support the findings of this study are available from the corresponding author upon reasonable request. 

\section{Acknowledgments}

The work at the College of William and Mary was sponsored by the DOE through Grant No. DEFG02-04ER46127. The work at the Department of Optical Science and Engineering, Fudan University, was supported by the National Natural Science Foundation of China with Grant No. 11774064, National Key Research and Development Program of China (Grant No. 2016YFA0300703), and National Key Basic Research Program (No. 2015CB921403). The work at the State Key Laboratory of Superlattices and Microstructures, Institute of Semiconductors, Chinese Academy of Sciences, was supported by National Natural Science Foundation of China with Grant No. U1632264.

\section{Author Contributions}
X. L., P. L., H. C. Y, J. H. Z., H. B. Z., and G. L. designed and analyzed the experiments. H. L. W., S. H. N., J. Y. S., and X. Z. Y. prepared the samples and carried out characterizations using MOKE, RHEED, and SQUID measurements. H. C. Y., J. Y. S., X. L., P. L., and F. J. performed the TRMOKE experiments. X. L., P. L., H. C. Y. and J. Y. S. conducted the data analysis and simulations. All authors discussed the results. X. L., P. L., J. H. Z., H. B. Z., and G. L. wrote the manuscript with contributions from all authors.

\section{Competing Interests}
The authors declare no competing financial interests.

\bibliographystyle{naturemag}
\bibliography{ms}

\end{document}